\pdfoutput=1
\documentclass{emulateapj}

\usepackage{amsmath,natbib,graphicx,euscript}

\def\cosmospp{{\texttt{Cosmos{\small++}}}\ }
\def\Ibar{{I\mkern-6.8mu\raise0.3ex\hbox{-}}}

\shorttitle{Bar-mode instability in magnetized neutron stars}
\shortauthors{Camarda et al.}
\slugcomment{To appear in ApJ}

\begin{document}

\title{Dynamical bar-mode instability in differentially rotating magnetized neutron stars} 

\author{Karen D. Camarda\altaffilmark{1}, 
     Peter Anninos\altaffilmark{2},
     P. Chris Fragile\altaffilmark{3} and
     Jos\'e A. Font\altaffilmark{4}}
\altaffiltext{1}{Department of Physics and Astronomy, 
              Washburn University, Topeka, KS 66621, USA}
\altaffiltext{2}{Lawrence Livermore National Laboratory, 
              P.O. Box 808, Livermore, CA 94550, USA}
\altaffiltext{3}{Department of Physics and Astronomy, 
              College of Charleston, Charleston, SC 29424, USA}
\altaffiltext{4}{Departamento de Astronom\'{\i}a y Astrof\'{\i}sica, 
              Universidad de Valencia, Dr. Moliner 50, 46100, 
              Burjassot, Valencia, Spain}

\begin{abstract}
This paper presents a numerical study over a wide 
parameter space of the likelihood of the dynamical bar-mode 
instability in differentially rotating magnetized neutron stars. 
The innovative aspect of this study is the incorporation of 
magnetic fields in such a context, which have thus far been 
neglected in the purely hydrodynamical simulations available 
in the literature. The investigation uses the \cosmospp code
which allows us to perform three-dimensional simulations
on a cylindrical grid at high resolution.
A sample of Newtonian magneto-hydrodynamical simulations starting from 
a set of models previously analyzed by other authors without 
magnetic fields has been performed, providing
estimates of the effects of magnetic fields on the dynamical
bar-mode deformation of rotating neutron stars. 
Overall, our results suggest that the effect of magnetic fields 
is not likely to be very significant in realistic configurations. 
Only in the most extreme cases are the magnetic fields able to 
suppress growth of the bar mode.
\end{abstract}



\section{Introduction}
\label{sec:intro}

Rotating neutron stars formed following the gravitational collapse of a massive stellar iron core or the accretion-induced collapse of a white dwarf can be subject to various nonaxisymmetric instabilities depending on the amount and degree of differential rotation. The prospects of detection of gravitational radiation from newly born rapidly rotating neutron stars by the current and future generations of kHz-frequency, ground-based gravitational wave interferometers highly motivate the investigation of such instabilities. In particular, if the rotation rate is high enough and shows a high degree of differentiation,  the star is subject to the so-called $m=2$, {\it dynamical} bar-mode instability driven by hydrodynamics and gravity, with $m$ being the order of the azimuthal nonaxisymmetric fluid mode $e^{\pm im\varphi}$ . On the other hand, at lower rotation rates gravitational radiation and viscosity can drive a star {\it secularly} unstable against bar-mode deformation. These two flavors of the bar-mode instability set in when the ratio $\beta=T/|W|$ of rotational kinetic energy $T$ to gravitational potential energy $W$ exceeds a critical value $\beta_{\rm c}$.   Early studies with incompressible MacLaurin spheroids in Newtonian gravity showed that the onset of the instability arises when $\beta_c\sim 0.27$ and $0.14$ for the dynamical and secular cases, respectively~\citep{chandra69}.

Improvements on these simplified analytic models have been achieved through numerical work. Newtonian and general relativistic analyses of the dynamical bar-mode instability  are available in the literature, using both simplified models based on equilibrium stellar configurations perturbed with suitable eigenfunctions, and more involved models for the core collapse scenario. 
Newtonian hydrodynamical simulations have shown that the value of $\beta_{\rm c}$ is quite independent of the stiffness of the equation of state, provided the star is not strongly differentially rotating (see~\citet{houser94,new00,liu02} and references therein). On the other hand, relativistic simulations~\citep{shibata00} have yielded a slightly smaller value of the dynamical instability parameter ($\beta_{\rm c}\sim 0.24-0.25$). They have also shown that the dynamics of the process closely resembles that found in Newtonian theory, that is, unstable models with large enough $\beta$ develop spiral arms following the formation of bars, ejecting mass and redistributing the angular momentum. Further relativistic simulations~\citep{baiotti07} have shown the appearance of nonlinear mode-coupling which can limit, and even suppress, the persistence of the bar-mode deformation.
It is also worth mentioning that, as the degree of differential rotation 
becomes higher and more extreme, 
Newtonian simulations~\citep{shibata02,shibata03} have also shown 
that rotating stars are dynamically unstable against bar-mode 
deformation even for values of $\beta$ of order 0.01.

Whether the requirements for the development of the instability inferred from numerical simulations are  met by the collapse progenitors remains unclear. Observations of surface velocities imply that a large fraction of progenitor cores are rapidly rotating. However, it has been shown that magnetic torques can spin down the core of the progenitor, leading to slowly rotating neutron stars at birth~\citep{spruit98}. The most recent computations of the evolution of massive stars, which include angular momentum redistribution by magnetic torques and spin estimates of neutron stars at birth, lead to core collapse progenitors which do not seem to rotate fast enough to guarantee the unambiguous growth of the canonical bar-mode instability~\citep{heger05,ott06}. These estimates are in agreement with observed periods of young neutron stars. However, rapidly-rotating cores might be produced by an appropriate mixture of high progenitor mass and low metallicity. Recent estimates suggest that about 1\% of all stars with masses larger than 10 solar masses will produce rapidly rotating cores~\citep{woosley06}. Recent relativistic simulations of a large sample of rotational core collapse models carried out by~\citet{dimmelmeier08}, which include a state-of-the-art treatment of the relevant physics of the collapse phase and realistic precollapse rotational profiles~\citep[see also][]{ott07}, have shown that the critical threshold of the dynamical bar-mode instability is never surpassed, at least early after core bounce. However, a large set of the models investigated by~\citet{ott07} and~\citet{dimmelmeier08} do show that models with sufficiently differential and rapid rotation are subject to the low$-\beta$ instability. In addition it is also worth mentioning the simulations of~\citet{baiotti08} which show that hypermassive differentially rotating neutron stars form following the merger of low-mass binary neutron stars, when modeled as polytropes. The resulting  hypermassive star undergoes a persistent phase of bar-mode oscillations, emitting large amounts of gravitational radiation prior to its delayed collapse to a black hole.

An important piece of physics that the existing numerical work has so far neglected is the presence of magnetic fields. The amplification and emergence of strong magnetic fields in neutron stars from initially weak magnetic field configurations in the pre-collapse stellar cores are currently under investigation~\citep[see e.g.,][and references therein]{burrows07,cerda-duran08}. We note, however, that the weakest point of all existing magneto-rotational core collapse simulations to date is the fact that both the strength and distribution of the initial magnetic field in the core are unknown. 

In this paper, we show results from a  detailed numerical study  of the effects that 
magnetic fields may have on the dynamical bar-mode instability in differentially 
rotating magnetized neutron stars. 
In particular, we investigate how sensitive the onset and development of the instability is to the presence of magnetic fields, as well as the role played by the magnetorotational instability (MRI) and magnetic braking mechanisms to alter the angular momentum distribution in the star and possibly suppress the bar-mode instability.  
Our study is motivated by the potential astrophysical implications that the presence of strong magnetic fields may have for post-bounce core collapse dynamics and, in turn, for gravitational wave astronomy. 

The uncertainty of the strength and distribution of magnetic fields in 
collapse progenitors is reflected in our somewhat ad hoc parameterization
of the field configuration through a large sample of equilibrium models
of rapidly and highly differentially rotating neutron stars.
Our sample of Newtonian magnetohydrodynamical (MHD) simulations is 
based upon the set of purely hydrodynamical models previously 
analyzed by~\citet{new00}. The simulations are performed  using the 
covariant (and adaptive mesh refinement) code 
\cosmospp~\citep{anninos03a,anninos03b,anninos05,fragile05} which allows 
us to perform three-dimensional simulations on a logarithmically
scaled cylindrical grid at high resolution.
A number of equilibrium models of rapidly rotating stars with different 
values of the rotational instability parameter ($\beta=T/|W|$) and magnetic plasma beta ($\beta_B=P/P_B$),
and different polytropic equations of state are constructed, introducing also 
different configurations for the magnetic field distribution (of both
poloidal and toroidal varieties) and field strengths. 
The equilibrium models are perturbed by seeding small random perturbations 
in order to initiate the onset of the bar-mode deformation.

The organization of the paper is as follows. Section~\ref{sec:methods} discusses
our basic formalism, numerical methods, diagnostics, and the construction
of initial data; Section~\ref{sec:results} presents our results in two 
subsections, one for initially toroidal field configurations and 
one for poloidal. We conclude with a summary and discussion of our 
results in Section~\ref{sec:conclusion}.

\section{Formalism, Initial Data, and Diagnostics}
\label{sec:methods}

\subsection{\cosmospp Framework}

Although the \cosmospp code now includes options to solve the full radiative and
conductive MHD equations for Newtonian systems plus multi-species chemical and nuclear 
reaction networks, we only include here the subset of those equations necessary 
for the current work. However, see \citet{anninos03b} for a more complete 
description of the Newtonian options including radiation, or \citet{anninos05} 
for the general relativistic formulation and a discussion of the
various energy formulation options available in the code. In the present study, we use
the internal energy (and artificial viscosity, $A.V.$) formulation due to
the robustness of the method in tracking adiabats across thin kinematically
dominated stellar atmospheres. The relevant equations are sufficiently different
than those published in our previous papers (simplified to Newtonian form
and generalized to covariant curvilinear grids) that we write them out
here for convenience:
\begin{eqnarray}
\frac{\partial (\sqrt{g}~\rho)}{\partial t} +\partial_i(\sqrt{g}~\rho~V^i)
     &=& 0 ~,
     \label{eqn:dens} \\
\frac{\partial (\sqrt{g}~S_k)}{\partial t} +\partial_i(\sqrt{g}~S_k~V^i)
     &=& \left(\frac{S^i S^j}{2\rho} - 
               \frac{B^i~B^j}{8\pi}\right) \sqrt{g}~\partial_k~g_{ij}  
               \nonumber \\
               &&+\frac{1}{4\pi} \partial_i(\sqrt{g}~B^i~B_k)
               \nonumber\\
     &&-\sqrt{g}~\partial_i \left[\left(P+P_B\right) \delta^i_k
               + Q^i_k\right] \nonumber \\
               &&- \sqrt{g}~\rho~\partial_k \Phi ~,
     \label{eqn:mom} \\
\frac{\partial (\sqrt{g}~e)}{\partial t} +\partial_i(\sqrt{g}~e~V^i)
     &=& -(P\delta^i_j + Q^i_j) ~\partial_i (\sqrt{g}~V^j) ~,
     \label{eqn:en} \\
\frac{\partial (\sqrt{g}~B^k)}{\partial t} +\partial_i (\sqrt{g}~B^k~V^i)
     &=& \sqrt{g}~B^i~\partial_i V^k
         - g^{ik}~\partial_i \psi ~,
     \label{eqn:mag} \\
\partial_i(\sqrt{g}~g^{ij}~\partial_j \Phi) &=& 4\pi G~\sqrt{g}~\rho~,
    \label{eqn:grav}
\end{eqnarray}
where $\partial_i \equiv \partial/\partial\xi^i$ represents covariant derivatives
in generalized coordinates $\xi^i$, and
$\sqrt{g}$ is the determinant of the spatial 3-metric $g_{ij}$
defining the coordinate system
(cylindrical for this work, with a logarithmically scaled radius 
to achieve greater resolution in the star's interior).
Also, $\rho$ is the fluid density,
$V^k$ is the contravariant fluid velocity, 
$S_k = \rho V_k$ is the covariant momentum, 
$e$ is the fluid internal energy density,
$Q^i_j$ is the artificial viscosity tensor (here we use
the unsplit finite volume version of the scalar viscosity 
from \citep{anninos05} with $k_q=2.0$ and $k_l=0.1$),
$P$ is the fluid pressure,
$B^k$ is the contravariant magnetic field vector,
$P_B = B^iB_i/8\pi$ is the magnetic pressure.
$\Phi$ is the gravitational potential, which is found by solving
equation (\ref{eqn:grav}) with multipole boundary conditions that include up to 15
spherical-polar harmonics and all corresponding azimuthal moments.
In this work, we assume an ideal gas equation of state
in the form $P=(\Gamma-1)e$.
The MHD equations are derived with the standard
assumptions relevant for many astrophysical problems:
the system is nonrelativistic and fully ionized, the displacement currents
in Maxwell's equations are neglected,
the net electric charge is small, and the characteristic
length scales are large compared to particle
gyroradii scales.

The scalar potential $\psi$ in the magnetic induction equation (\ref{eqn:mag})
is introduced as a divergence cleanser to maintain a divergence-free
magnetic field ($\partial_i(\sqrt{g}~B^i) = 0$).  Options are included
in Cosmos to
solve any one of the following constraint equations for $\psi$ \citep{dedner02}: 
\begin{eqnarray}
\nabla^2 \psi &=& - \frac{\partial[\partial_i(\sqrt{g}~B^i)]}{\partial t}
            \approx  - \frac{\partial_i(\sqrt{g}~B^i)}{\Delta t} , 
            \label{eqn:psi_ell} \\
\psi &=& - c_p^2 \partial_i (\sqrt{g} B^i) , \\
\frac{\partial \psi}{\partial t} &=& - \frac{c_h^2}{c_p^2} \psi
         - c_h^2 \partial_i \left(\sqrt{g} B^i\right) ~,
\label{eqn:psi}
\end{eqnarray}
which correspond, respectively, to elliptic, parabolic, and
mixed hyperbolic and parabolic constraints.
Here $c_p$ and $c_h$ are user-specified constants used to
regulate the filtering process and weight the relative
significance of the hyperbolic and parabolic components.
Also the time derivative in equation (\ref{eqn:psi_ell}) is approximated
as a finite difference with zero divergence at the initial time.
For all of the calculations presented in this paper, we use
the mixed hyperbolic and parabolic form with parameters 
$c_h = 0.2 c_\mathrm{cfl} \Delta x_\mathrm{min}/\Delta t$ and $c_p^2 = 0.3 c_h$, 
where $c_\mathrm{cfl}=0.4$ is the Courant coefficient, 
$\Delta x_\mathrm{min}$ is the minimum covariant zone length, 
and $\Delta t$ is the evolution time step.

Equations (\ref{eqn:dens}) -- (\ref{eqn:grav}) and (\ref{eqn:psi}) are solved
in a modified cylindrical coordinate system 
$\xi^i = (\eta,z,\phi)$, where $\eta=\ln (\varpi+1)$ is a logarithmic radial coordinate 
used to concentrate resolution toward the interior of the star, 
with $\varpi=r\sin \theta$ being the usual cylindrical radius. 
With this coordinate choice, the line element for the metric $g_{ij}$ becomes
\begin{equation}
dl^2 = e^{2\eta}~d\eta^2 + dz^2 + (e^\eta -1)^2~d\phi^2 ~,
\end{equation}
and $\sqrt{g} = e^\eta(e^\eta - 1)$.
We consider two different mesh
resolutions, $64^3$ and $96^3$, to address the robustness and relative
convergence of our results. These are about optimal
resolutions for three-dimensional simulations where a large number of 
parameters are to be explored, particularly with active magnetic fields
which increase substantially the computational workload and suffer
greater wave speed restrictions on the time step.

\subsection{Initial Data}

\subsubsection{Rotating Polytropes}

We begin by constructing equilibrium models of rapidly rotating polytropic 
stars using Hachisu's self-consistent field technique \citep{hachisu86}. 
For an initially axisymmetric configuration with angular velocity 
$\Omega =\Omega (\varpi)$ that depends only on the distance of the fluid from 
the rotation axis ($\varpi=r \sin\theta$), the equilibrium configuration satisfies 
\begin{equation}
\Phi + h_0^2 \Psi + H = C_0 ~,
\label{eqn:equilib}
\end{equation}
where $h_0$ and $C_0$ are constants,
$H=\int \rho^{-1} dP$ is the fluid enthalpy,
$\Phi$ is the gravitational potential obtained by solving the Poisson equation
(\ref{eqn:grav}), and
\begin{equation}
\Psi(\varpi) = -\frac{1}{h_0^2} \int \Omega^2(\varpi) \varpi d\varpi
\label{eqn:rotate}
\end{equation}
describes the rotational profile of the star. 
For a polytropic gas with $P=\kappa \rho^\Gamma = \kappa \rho^{(1+1/N)}$, we have
\begin{equation}
H=(N+1)\kappa \rho^{1/N} = \left( \frac{H_\mathrm{max}}{\rho_\mathrm{max,0}^{1/N}} \right) 
                        \rho^{1/N} ~,
\label{eq:enthalpy}
\end{equation}
where $\rho_\mathrm{max,0}$ is the initial maximum density.

A variety of rotation profiles can be considered, such as rigid rotation 
($\Omega = \mathrm{const.}$), constant linear velocity ($\Omega = V_0/\varpi$), 
or constant specific angular momentum ($\Omega = j_0/\varpi^2$, where $j$ is the 
specific angular momentum of the fluid). In our case, we choose a Maclaurin 
spheroid profile to compare with the earlier work of unmagnetized neutron stars
\citep{new00}
\begin{equation}
\Omega(\varpi) = h_0 \left[ 1 - \left( 1 - \frac{m(\varpi)}{M} \right)^{2/3} \right] 
              \varpi^{-2} ~,
\end{equation}
where $M$ equals the total mass of the star (spheroid), 
$m(\varpi)$ is the mass interior to $\varpi$, and 
\begin{equation}
h_0= \frac{5J}{2M}
\end{equation}
can now be specified in terms of the total angular momentum $J$ of the star.

The constants $h_0$ and $C_0$ in equation (\ref{eqn:equilib}) are set by an 
appropriate choice of boundary conditions. Specifically, we require that 
$\rho$, $P$, and $H$ vanish at the surface of the star. Any two points on the 
surface of the star can then be used to specify a solution; we choose a point on the 
equator (point $A$) and one of the poles (point $B$) by specifying the equatorial 
surface radius $\varpi_E$ and axis ratio $z_P/\varpi_E$, where $z_P$ is the polar radius. 
The constants are then given as
\begin{equation}
h_0^2 = -\frac{\Phi(A)-\Phi(B)}{\Psi(A)-\Psi(B)}
\end{equation}
and 
\begin{equation}
C_0 = \Phi(A) + h_0^2 \Psi(A) ~.
\end{equation}
Finally we must specify $\Gamma$ (or $N$) and $\kappa$ (or $\rho_\mathrm{max,0}$) 
to close the system of equations. The solution proceeds by guessing an initial 
distribution for $\rho$; solving equations (\ref{eqn:grav}) and 
(\ref{eqn:rotate}) for $\Phi$ and $\Psi$, respectively; using equation 
(\ref{eqn:equilib}) to set $H$ and $H_\mathrm{max}$; and then using 
equation (\ref{eq:enthalpy}) to determine a new density distribution. 
This procedure is repeated iteratively until the solution converges 
sufficiently (for us, once $\Delta h_0/h_0$, $\Delta C_0/C_0$, and 
$\Delta H/H$ are all $\le 10^{-4}$).

The initial data are solved
in two-dimensional logarithmic cylindrical coordinates using the same
polar and radial resolutions as the full three-dimensional grid that we use for the production
simulations, which we then map (i.e., rotate) azimuthally
onto the three-dimensional grid. The quality of solutions is verified by computing the virial error
(V.E.) as
\begin{equation}
\mathrm{V.E.} = 2T + W + 3\int P dV ~,
\end{equation}
where $T$ is the total kinetic energy, $W$ is the total gravitational
potential energy, $P$ is the thermal pressure and $dV$ is the volume element.
Virial errors in our calculations are small:
$4\times10^{-4}$ for the $\Gamma=5/3$ cases,
$8\times10^{-4}$ for $\Gamma=2$, and
$1.6\times10^{-3}$ for $\Gamma=3$. Normalized to the absolute value of
the potential energy, we find $|\mathrm{V.E.}/W| \lesssim 3\times10^{-3}$ for all cases.

Once the equilibrium density distribution $\rho_\mathrm{EQ}$ is determined, 
we apply a small random perturbation of the form 
$\rho(R,z,\phi) = \rho_\mathrm{EQ}[1+a_0 f(R,z,\phi)]$, 
where $a_0=10^{-2}$ is the perturbation amplitude and $f$ is a random number 
between -1 and 1. This perturbation serves as a seed for the bar-mode 
and magneto-rotational instabilities to grow.

Because MHD codes have difficulty treating pure vacuums, 
it is necessary to initially fill the regions of the grid outside 
the star with a low density ($\rho_\mathrm{floor} = 10^{-6} \rho_\mathrm{max,0}$), 
low energy ($e_\mathrm{floor} = \kappa \rho_\mathrm{floor}^\Gamma/(\Gamma-1)$), 
static ($V^i=0$) background. During the evolution, any time the fluid 
density or energy attempt to drop below their respective floor values they are reset.

\subsubsection{Magnetic Fields}

In this work, we choose two idealized initial configurations for the magnetic fields, one purely toroidal and one purely poloidal. In all cases the fields are chosen to be initially weak, so that in some sense they simply represent additional perturbations away from the initial equilibrium state. However, it is now well understood that differentially rotating fluids with weak magnetic fields are susceptible to the MRI \citep{balbus91}. The only criterion required to trigger the MRI is
\begin{equation}
\frac{d\Omega^2}{d \ln \varpi} < 0 ~,
\label{eqn:mricondition}
\end{equation}
a condition which is clearly met inside our model star. 
Thus, our initially weak magnetic fields can potentially become dynamically 
important during the evolution (provided we have sufficient resolution to capture the MRI). 

The toroidal field case begins with $B^\eta=B^z=0$ and 
\begin{equation}
B^\phi = \text{max}\left[ 0, \quad C_T \frac{e^{-\Delta\xi^2/(2\sigma^2)} - e^{-0.5}}
                                         {\sqrt{2\pi}\sigma} \right] ~,
\end{equation}
where $\Delta\xi = \vert \xi^i - \xi^i_\mathrm{loop} \vert$ is the position
offset relative to the location of the central strongest field loop
$\xi^i_\mathrm{loop} = (\eta_\mathrm{loop}, 0, \phi)$. The characteristic
half-width of the loop is defined as $\sigma = \varpi_E/8$, and
$\eta_\mathrm{loop}$ is set to either $\varpi_E/2$ or $\varpi_E/4$ so that the center
of the toroidal loop is initially located at radius $r_\mathrm{loop} \approx 0.65 \varpi_E$
or $r_\mathrm{loop} \approx 0.28 \varpi_E$.
The normalization constant $C_T$ is chosen to satisfy our particular 
choice for $\beta_\mathrm{B,min}$, the initially lowest value for 
$P/P_B$ inside the star. The extra factor of $e^{-0.5}$ is included to 
ensure that the magnetic field is initially confined to the interior of the star.

The poloidal field case begins with $B^\phi=0$. 
The other two magnetic field components are specified from a magnetic vector
potential using the following divergence-free construction:
\begin{equation}
\sqrt{g} B^i = \left[-\frac{\partial}{\partial z}   \left(\sqrt{g_{\phi\phi}} A_\phi\right),
             \quad \frac{\partial}{\partial \eta}\left(\sqrt{g_{\phi\phi}} A_\phi\right),
             \quad 0 \right] ~.
\end{equation}
For the vector potential we use the solution corresponding to a current loop 
in the equatorial plane of radius $r_\mathrm{loop}$ (which we typically
set to $\varpi_E/2$) centered on the origin
\citep{jackson75}
\begin{equation}
A_\phi = C_P \sum_{n=0}^\infty \frac{(-1)^n(2n-1)!!}{2^n(n+1)!}
                            \frac{r_<^{2n+1}}{r_>^{2n+2}} P^1_{2n+1}(\cos \theta) ~,
\end{equation}
where $(2n-1)!! = (2n-1)(2n-3)(...)(5)(3)(1)$, 
$r_< = \mathrm{min}(r, r_\mathrm{loop})$, 
$r_> = \mathrm{max}(r, r_\mathrm{loop})$, and $P^m_l$ are the Legendre polynomials.
In practice, we calculate the first 20 terms of the sum. 
To prevent the magnetic field from extending beyond the surface of the star,
$A_\phi$ is truncated at values smaller than $0.5 A_{\phi, \mathrm{max}}$,
where $A_{\phi, \mathrm{max}}$ is the maximum value of $A_\phi$.
Again, the normalization constant $C_P$ is chosen to satisfy our choice of 
$\beta_\mathrm{B,min}$.

As a rough guideline to determine whether there is
sufficient numerical resolution to capture the MRI,
we compare the grid cell spacing at the position where $\beta_B=\beta_{B,\text{min}}$
against the characteristic (minimum unstable) MRI wavelength 
\begin{equation}
\lambda_\mathrm{MRI} = \frac{2\pi v_A}{\Omega} ~,
\end{equation}
where $v_A^2 = 2 P_B/\rho$ is the Alfv\'en velocity of the plasma.
The ratio of $\lambda_\mathrm{MRI}$ over the grid spacing $\Delta \varpi$ at
$\xi^i(\beta_B=\beta_{B,\text{min}})$ is given 
in Table \ref{tab:params}, along with the maximum field amplitude
inside the star at the initial time.

\subsubsection{Grid, Units, and Parameters}

In all calculations the radial box size is set to three times the equatorial radius
of the star $L_\varpi = 3\varpi_E$. The box size along the $z$-axis
is set smaller than the radial dimension $L_z = 1.2 \varpi_E$, compensating
for the high aspect ratio of the stellar profile while also maintaining reasonable
resolution along the $z$-axis and accurate multipole boundary conditions
for the gravity solver.
The logarithmic nature of the coordinate system results in a spatial
resolution of $\Delta\varpi \approx 0.022 \varpi_E$ and $0.014 \varpi_E$ near the origin
along the radial axis for the $64^3$ and $96^3$ grids respectively.
This is more than a factor of two improvement in resolution  over a uniform grid,
and increases the radial resolution to nearly match the polar
($\Delta z \approx 0.01875 \varpi_E$
and $0.0125 \varpi_E$ for the $64^3$ and $96^3$ grids).
For reference, the azimuthal resolution in the equatorial plane
of the $64^3$ grid varies from $\varpi\Delta\phi \approx 0.05\varpi_E$ 
at $\varpi \sim \varpi_E/2$ to about $10^{-3}\varpi_E$ near the origin.

Displayed results in all subsequent figures
are presented in dimensionless code units.
We define the equatorial equilibrium surface radius to be the unit coordinate dimension
($\tilde{\ell} = \varpi_E= 1.7\times10^8~\text{cm}$), the dynamical time of a spheroid
with equatorial surface radius $\varpi_E$ to be the unit interval of time
($\tilde{t} = \sqrt{\varpi_E^3/GM} \approx 0.1$ s for the $\Gamma=5/3$ runs), 
and the maximum central density of the star 
to be the unit density ($\tilde{\rho} = \rho_{\text{max,0}}=10^{10}~\text{g/cm}^3$).
This in turn implies a scaled magnetic field unit of
$\sqrt{G M \rho_{\text{max,0}}/\varpi_E}$ ($\approx 1.7\times10^{14}$ Gauss for
$\Gamma=5/3$).

Table \ref{tab:params} lists all of the numerical simulations we have performed in this
study, together with their corresponding parameter sets. The calculations represent
a large parameter space including the field strength
characterized by the minimum plasma beta $\beta_\mathrm{B,min}$, the field orientation
(toroidal or poloidal), the equatorial field or current loop radius $r_\mathrm{loop}$,
the equation of state polytropic index $\Gamma$, 
the rotational instability parameter $\beta$,
the polar to equatorial axis ratio $A_r$,
and the grid resolution. Also shown in Table \ref{tab:params} is the maximum
magnetic field amplitude inside the star, and the ratio of the MRI wavelength
over the local radial cell resolution.
The prefixes ``T'' and ``P'' in our naming convention refer to initial toroidal 
and poloidal field orientations, respectively. Runs labeled without either of
these prefixes are the baseline unmagnetized calculations.

\begin{deluxetable*}{lcccrcccc}
\tablecolumns{7}
\tablewidth{0pt}
\tabletypesize{\scriptsize}
\tablecaption{Simulation Parameters
\label{tab:params}}
\tablehead{
\colhead{Simulation} & \colhead{Resolution} & \colhead{$\beta_\mathrm{B,min}$\tablenotemark{a}} &
\colhead{$\beta$\tablenotemark{b}}  & \colhead{$\Gamma$ ($N$)\tablenotemark{c}}   & \colhead{$A_r$\tablenotemark{d}}    &
\colhead{$r_\mathrm{loop}$\tablenotemark{e}} &
\colhead{$\lambda_\mathrm{MRI}/\Delta \varpi$\tablenotemark{f}} &
\colhead{$|B|_{\text{max}}$\tablenotemark{g}}
}
\startdata
G53Binf   & $64^3$  & $\infty$ & 0.295 & 5/3 (1.5) & 0.208 & --- & ---  & -- \\
G53BinfHR & $96^3$  & $\infty$ & 0.295 & 5/3 (1.5) & 0.208 & --- & ---  & -- \\
G20Binf   & $64^3$  & $\infty$ & 0.286 & 2 (1.0)   & 0.250 & --- & ---  & -- \\
G30Binf   & $64^3$  & $\infty$ & 0.298 & 3 (0.5)   & 0.250 & --- & ---  & -- \\
\\
TG53B1     & $64^3$  & 1        & 0.295 & 5/3 (1.5) & 0.208 & 0.65 & 35.0 & $5.4\times10^{13}$  \\
TG53B10    & $64^3$  & 10       & 0.295 & 5/3 (1.5) & 0.208 & 0.65 & 11.0 & $1.7\times10^{13}$  \\
TG53B100   & $64^3$  & 100      & 0.295 & 5/3 (1.5) & 0.208 & 0.65 & 3.5  & $5.4\times10^{12}$  \\
TG53B100R3 & $64^3$  & 100      & 0.295 & 5/3 (1.5) & 0.208 & 0.28 & 5.6  & $2.9\times10^{13}$  \\
TG53B100HR & $96^3$  & 100      & 0.292 & 5/3 (1.5) & 0.208 & 0.65 & 5.6  & $5.5\times10^{12}$  \\
TG53B500   & $64^3$  & 500      & 0.295 & 5/3 (1.5) & 0.208 & 0.65 & 1.6  & $2.4\times10^{12}$  \\
TG53B500R3 & $64^3$  & 500      & 0.295 & 5/3 (1.5) & 0.208 & 0.28 & 2.7  & $1.4\times10^{13}$  \\
TG53B1e8   & $64^3$  & 1e8      & 0.295 & 5/3 (1.5) & 0.208 & 0.65 & 0.003& $5.4\times10^{9}$  \\
TG20B100   & $64^3$  & 100      & 0.286 & 2 (1.0)   & 0.250 & 0.65 & 4.4  & $1.8\times10^{13}$  \\
TG30B100   & $64^3$  & 100      & 0.298 & 3 (0.5)   & 0.250 & 0.65 & 5.2  & $4.1\times10^{13}$  \\
\\
PG53B10    & $64^3$  & 10       & 0.295 & 5/3 (1.5) & 0.208 & 0.5  & 12.8 & $2.3\times10^{13}$  \\
PG53B10HR  & $96^3$  & 10       & 0.292 & 5/3 (1.5) & 0.208 & 0.5  & 20.5 & $2.7\times10^{13}$  \\
PG53B100   & $64^3$  & 100      & 0.295 & 5/3 (1.5) & 0.208 & 0.5  & 4.1  & $7.2\times10^{12}$  \\
PG53B100HR & $96^3$  & 100      & 0.292 & 5/3 (1.5) & 0.208 & 0.5  & 6.5  & $8.5\times10^{12}$  \\
PG53B500   & $64^3$  & 500      & 0.295 & 5/3 (1.5) & 0.208 & 0.5  & 1.8  & $3.2\times10^{12}$  \\
PG53B500HR & $96^3$  & 500      & 0.292 & 5/3 (1.5) & 0.208 & 0.5  & 2.9  & $3.8\times10^{12}$  \\
PG30B100   & $64^3$  & 100      & 0.298 & 3 (0.5)   & 0.250 & 0.65 & 6.0  & $4.2\times10^{13}$  \\
\enddata

\tablenotetext{a}{$\beta_{B,\text{min}}=(P/P_B)_\text{min}$ 
is the plasma parameter defining the minimum
hydrodynamic to magnetic pressure ratio (or maximum magnetic field amplitude).}
\tablenotetext{b}{$\beta=T/|W|$ is the rotational instability parameter.}
\tablenotetext{c}{$\Gamma =1+1/N$ is the adiabatic index for the ideal gas equation of state.}
\tablenotetext{d}{$A_r$ is the polar to equatorial axis ratio of the initial star configuration.}
\tablenotetext{e}{$r_\mathrm{loop}$ is the radius of the field or current loop in the equatorial
plane.}
\tablenotetext{f}{$\lambda_\mathrm{MRI}/\Delta \varpi$ is the ratio of the MRI wavelength to the
(cylindrical) radial cell width ($\Delta \varpi$) at a position corresponding
to the minimum initial plasma beta ($\beta_{B,\text{min}}$) inside the star.}
\tablenotetext{g}{$|B|_{\text{max}}$ is the maximum magnetic field amplitude inside the star
in units of Gauss.}

\end{deluxetable*}

\subsection{Gravitational Radiation Diagnostic}
\label{subsec:gwaveform}

Because the spacetime metric is held fixed in our simulations, 
we compute the gravitational radiation produced by these systems 
in the (traceless) quadrupole approximation. For an observer located on the 
symmetry axis, the two polarizations of the
gravitational wave amplitude are written as
\begin{eqnarray}
h_+ &=& \frac{G}{c^4} \frac{1}{r} \left(\skew6\ddot{\Ibar}_{xx} - \skew6\ddot{\Ibar}_{yy}\right) ~, \\
h_\times &=& \frac{G}{c^4} \frac{2}{r} \skew6\ddot{\Ibar}_{xy} ~,
\end{eqnarray}
where $\ddot{\Ibar}_{ij}$ is second time derivative of the reduced quadrupole 
moment of the mass distribution, given in flat space Cartesian coordinates $x^k$ by
\begin{equation}
\Ibar_{ij} = \int \rho (x_i x_j - \frac13 \delta_{ij} r^2) \, d^3 x ~,
\label{eq:Ibar}
\end{equation}
and $r=\sqrt{x^2 + y^2 + z^2}$ is the spherical distance to the source center of mass.
The integral is evaluated over the entire mass distribution.

It is well known that the
straight-forward procedure of computing the components of $\Ibar_{ij}$ 
at each time step, and then taking the needed time derivatives of the 
result numerically, produces an unacceptable level of numerical noise 
in the resulting waveforms \citep{Finn90}.
Instead we differentiate equation~(\ref{eq:Ibar}) with respect to time twice, 
each time using the evolution equations~(\ref{eqn:dens}) -- (\ref{eqn:mag})
to replace the time derivatives that appear 
in the resulting integrand. Many of the spatial derivatives which are 
introduced by this procedure can then be eliminated using integration 
by parts. The resulting expression for $\skew6\ddot{\Ibar}_{ij}$ is similar to
that given in \citet{new00}, except here we include a potential term missing 
from that expression, and add artificial viscosity and magnetic field contributions:
\begin{eqnarray}
\skew6\ddot{\Ibar}_{ij}
&=& \int \left[  2\rho V_i V_j - \frac23 \delta_{ij} \rho V^k V_k \right.\nonumber \\
               &&\phantom{\int}\left.- \rho x_j \partial_i \Phi - \rho x_i \partial_j \Phi  + \frac23 \delta_{ij} \rho x^k \partial_k \Phi \right.
    \nonumber \\
&&  \phantom{\int} \left.
    - \frac{1}{4\pi} \left(2 B_i B_j -  \delta_{ij} B^k B_k \right) 
                  - \frac{2}{3} \delta_{ij} \left(\frac{B^k B_k}{8\pi}\right)\right. \nonumber \\
    &&\phantom{\int}\left. + Q_{ij} + Q_{ji} - \frac23 \delta_{ij} Q^k_k \right] \, d^3 x ~.
\end{eqnarray}
In the following sections, we present the quantity $rh_+$,
to scale out the radial dependence, and normalize the wave amplitudes
by $(GM/\varpi_E c^2)^2$ so  that a direct comparison can be made to previous
unmagnetized calculations.

\section{Results}
\label{sec:results}

\subsection{Toroidal Magnetic Field Configurations}
\label{subsec:torresults}

\begin{figure}
\plotone{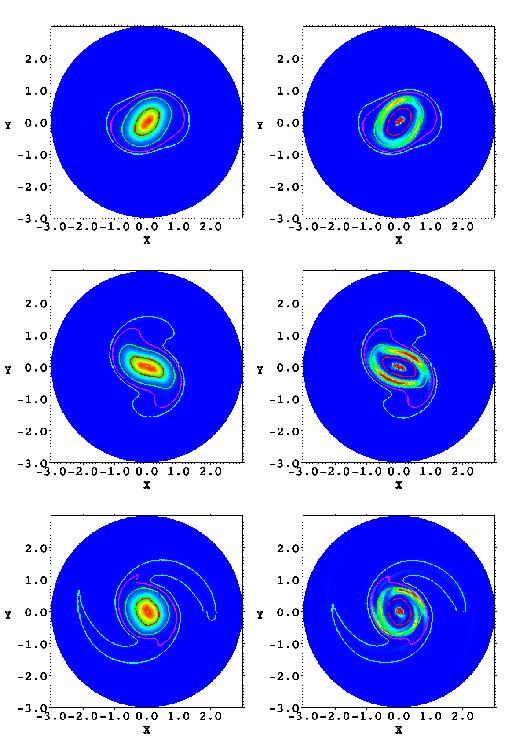}
\caption{Development of the bar mode
in the $\Gamma = 5/3$ simulations with a magnetic field (TG53B100, right column)
and without (G53Binf, left column).
Rows represent snapshot solutions at times: $t=$ 12, 14, and 20
in dynamical (code) units ($\tilde{t} = \sqrt{\varpi_E^3/GM}$).
Contour levels represent mass density in the equatorial plane at four
values normalized to the initial maximum density $\rho_{\text{max,0}}$:
(0.5, 0.05, 0.005, and 0.0009).
Linear color maps represent mass density in the case without magnetic fields
(left column), and magnetic pressure in the case with fields (right).
The magnetic pressure color scale is fixed for all the images to
cover the range $0 \rightarrow 10^{-4}$. The mass density color scale 
is adjusted to the maximum density at each time, but does not
deviate much from $0 \rightarrow 1.07$ over all displayed sequences.
\label{fig:images}}
\end{figure}

All calculations with toroidal configurations begin with an initial axisymmetric
solution in near equilibrium, and subsequently evolve in a similar fashion. They
display the characteristic exponential growth of the
even (primarily $m=2$ and 4) non-axisymmetric modes at early times,
followed by a mode saturation phase during
which the star develops an elongated
bar structure with spiral arms, then a final bar attenuation
phase that redistributes angular momentum as the
star evolves into a more axisymmetric configuration again. The early and
intermediate phases of the bar-mode growth are illustrated in Figure \ref{fig:images}
where we show images of the mass density (for the
G53Binf case) and magnetic
pressure (for the TG53B100 case) in the equatorial plane. 
Both sets of images display the same
contour levels of the mass density (0.5, 0.05, 0.005, 0.0009) normalized to
the initial maximum mass density $\rho_{\text{max,0}}$.
The mass densities are virtually identical in both magnetized and unmagnetized runs.
The magnetic pressure is initially confined within density contour levels 0.05 and 0.5,
a region about $2\sigma = 0.25\varpi_E$ thick in radius.
The magnetic field for this configuration is not buoyant
but remains mostly confined to those level boundaries as the bar mode takes shape.

\begin{figure}
\plotone{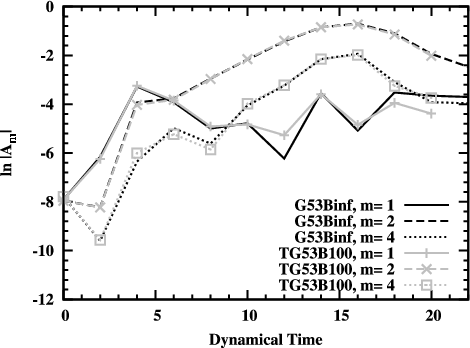}
\caption{Growth of the first few azimuthal Fourier mode amplitudes 
$|A_m|$ in the mass density, comparing two $\Gamma=5/3$ cases:
unmagnetized (G53Binf, dark line types)
and magnetized (TG53B100, gray lines and symbols). 
Results are derived by averaging the density inside a ring centered at 
$\varpi=0.45\varpi_E$ in the equatorial plane, extending one zone
deep in both radial and polar directions.
We do not plot the $m=3$ curve here for reasons of clarity, but note
that it roughly tracks the $m=1$ curve in shape and amplitude.
\label{fig:mode53}}
\end{figure}

Comparing mass density contours from the two simulations
in Figure \ref{fig:images}, it is apparent that the introduction
of magnetic fields (at least for this particular field
configuration) does not affect appreciably the growth of the bar mode.
This is confirmed in Figure \ref{fig:mode53} where we compare the growth
of the amplitude $|A_m|$ for the first few nonaxisymmetric modes ($m=$ 1, 2, and 4)
in the same two runs as Figure \ref{fig:images} (G53Binf and TG53B100). 
For reasons of clarity, we do not
show the $m=3$ curve, but note instead that the overall shape and amplitude
are very similar to the $m=1$ mode.
The quantity $|A_m|$ plotted in Figure \ref{fig:mode53} is the same as that
introduced in \citet{new00}. It represents the mode amplitude inside a ring centered 
at a fixed cylindrical radius ($\varpi=0.45\varpi_E$) in the equatorial plane
($z=0$)
\begin{equation}
|A_m| = \frac{|C_m|}{|C_0|}
   = \frac{\int_0^{2\pi} \overline{\rho}~e^{-im\phi}~d\phi}
          {\int_0^{2\pi} \overline{\rho}~d\phi} ~,
\end{equation}
where $\overline{\rho}$ is the density in the ring, averaged over cells along the
radial and polar axes as specified by the thickness of the ring in those directions.

\begin{figure}
\plotone{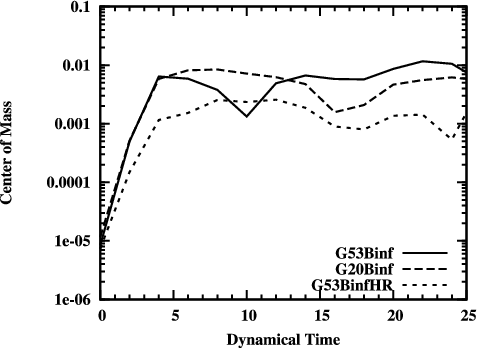}
\caption{Spherical radial position of the center of mass in a few
typical unmagnetized runs.
The vertical axis is dimensionally normalized by the initial equatorial
surface radius $\varpi_E$. Apart from a brief drift at early times,
the center of mass movement quickly saturates and remains confined
to a region that is within a half/quarter cell width radius
for the $64^3$/$96^3$ grids.
\label{fig:com}}
\end{figure}

An obvious result taken from Figure \ref{fig:mode53} is the strong
similarity in the early exponential growth and late time saturation profiles
of both magnetized and unmagnetized runs. But there are also several other
points of interest in Figure \ref{fig:mode53}. For example, the $m=1$
mode saturates after about five dynamical times to a level that
is well below those of the $m=2$ and 4 modes (about three orders of magnitude 
below $m=2$, and two orders below $m=4$ at their peaks), as is characteristic
of the bar mode. Since the odd modes tend to encapsulate numerical errors
such as the center of mass drift and loss of angular momentum, it is
encouraging to see the odd modes saturate in time and the characteristic
even modes to dominate the Fourier spectrum in convincing fashion.
We have tracked the mass center in our calculations and plot the result
for a few typical cases in Figure \ref{fig:com}. The early drift in the center of mass
quickly saturates within a few dynamical times, and the center of mass thereafter
remains mostly stationary. The total late-time drift in the position of the star
mass center is very small, $\sim 0.01\varpi_E$ for the $64^3$ grids, effectively
confined to less than half a cell width in the central most highly resolved
(smallest zoning) portion of the grid. The center of mass position is preserved
even better in the high-resolution $96^3$ runs, being confined to
within one quarter of a cell width, $\sim 0.0025\varpi_E$.

Although we do not show the results here, we have verified that the modal histories
in the magnetized runs are very similar to the unmagnetized results also for the
$\Gamma=$ 2 and 3 cases, and both sets look like the $\Gamma=5/3$ results
in Figure \ref{fig:mode53}.
The only major difference that we have observed and attributed to the
equation of state is a systematic shift (or delay) in time required
for the even bar modes to enter the exponential growth phase.
This is discussed further in the paragraphs below in the context of
gravitational wave emissions.

\begin{figure}
\plotone{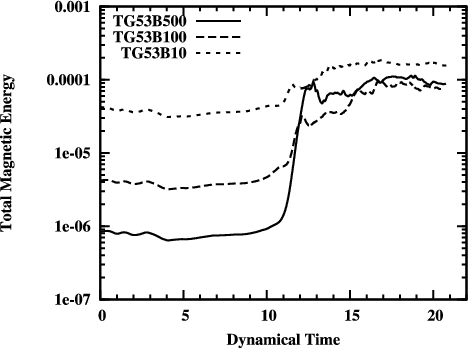}
\caption{Total magnetic energy integrated over the entire grid
for a few of the $\Gamma=5/3$ cases.
The horizontal axis is plotted in dynamical time units
$\sqrt{\varpi_E^3/GM}$, and the field energy in dimensionless code units.
\label{fig:fieldenergy}}
\end{figure}

\begin{figure}
\plotone{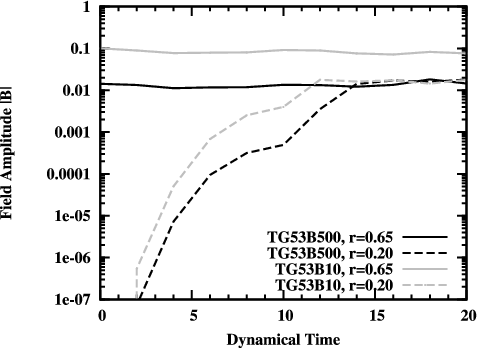}
\caption{Magnetic field amplitude averaged azimuthally inside a circular
annulus in the equatorial plane between radii
$0.65\varpi_E$ and $0.2\varpi_E$.
The horizontal axis is plotted in dynamical time units
$\sqrt{\varpi_E^3/GM}$, and the field amplitude in dimensionless code units
of $\sqrt{G M \rho_{\text{max,0}}/\varpi_E} \approx 1.7\times10^{14}$ Gauss.
\label{fig:shellamp}}
\end{figure}

The ineffectiveness of
toroidal magnetic fields is demonstrated again by two
additional calculations in which we varied $\beta_{B,\text{min}}$
to generate different initial field amplitudes:
a factor of five reduction in run TG53B500, and a ten-fold increase
in run TG53B10. In both cases, toroidal fields do
not affect significantly the growth or development of the bar mode,
in spite of the field amplification observed
in Figures \ref{fig:fieldenergy} and \ref{fig:shellamp} which plot
the total integrated magnetic energy and the local field amplitude, respectively, as a
function of time. The field amplitude in Figure \ref{fig:shellamp}
is calculated as an azimuthal average inside a circular annulus
in the equatorial plane at radii 0.65$\varpi_E$ (the
initial center of the toroidal field loop) and 0.2$\varpi_E$.
Note the field amplitude is already saturated and does not evolve much at the initial 
configuration radius ($r_\mathrm{loop}=0.65\varpi_E$), but grows substantially
at smaller radii. Eventually, the field amplitude and energy saturate in time at
all radii to roughly the same value.
This behavior is seen in all of the cases we have tried, including the
smallest field amplitude case (TG53B1e8) in which the field grows from
$|B| \sim 3\times10^{-5}$ to $\sim 0.02$ at radius 0.65$\varpi_E$ by runs end.
A confirmation of the convergence of these results is provided by the high-resolution
run TG53B100HR which falls between the B10 and B500 results, and tracks closely
the corresponding energy and field amplitudes at lower resolution.
The only difference between the high and lower resolution curves
is the growth phase in the high resolution case is triggered about one dynamical
time earlier (approximately a 10\% shift in time),
but saturates at the same mean field amplitude.

The exponential growth of the magnetic energy around a time $t \sim 11$ suggests 
the onset of a powerful magnetic field amplification mechanism. The delayed onset 
and steep exponential growth particularly favor axisymmetric modes of the MRI. 
These modes have the shortest growth times, but require a poloidal field component 
to act upon. For the initially toroidal configurations, significant poloidal fields 
are not present until the bar mode begins redistributing material within the star, 
explaining the delayed onset. Furthermore, it is clear from
Figures \ref{fig:fieldenergy} and \ref{fig:shellamp} that the magnetic field saturates
to roughly the same level at all radii and for all cases, regardless of initial
amplitudes. The saturation level appears comparable to the initial field
amplitudes in the B500 case: field amplitudes in cases initially greater than B500
are dissipated through hydrodynamic processes; field amplitudes in cases initially
less than B500 are amplified until they reach B500 levels before leveling off.
Our calculations suggest no mechanism exists which can drive field amplitudes
above B500 levels, implying that we are safely modeling the upper limit
of self-generated field strengths.

\begin{figure}
\plotone{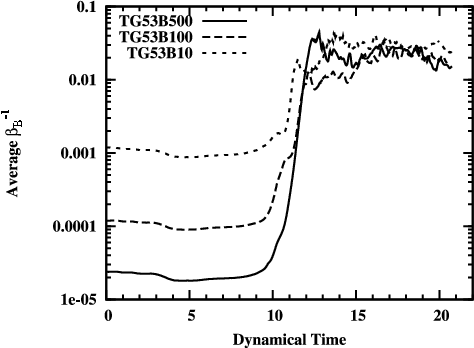}
\caption{Evolution of the mass density weighted average of the inverse
of the magnetic plasma beta ($\beta_B^{-1}$) inside the star
(satisfied by $\rho/\rho_{\text{max}} > 10^{-4}$, where $\rho_{\text{max}}$ is the
maximum density in the star at any given time). The high-resolution
run TG53B100HR (not shown here for reasons of clarity) closely resembles 
the growth curve from the corresponding lower resolution case TG53B100.
\label{fig:betainv53}}
\end{figure}

Although field amplification does indeed take place in all magnetized
runs we have performed (medium and high resolution), it falls well
short of thermal equipartition so it cannot easily affect
the dynamical evolution of the star. This is demonstrated
in Figure \ref{fig:betainv53} which shows the mass density weighted average
of the inverse plasma beta ($1/\beta_B$) inside the star.
In all cases the increase of magnetic pressure
saturates at a level that is significantly less than
10\% of the thermal pressure averaged across the star
(with maximum peak values of about 4\% for the $64^3$ runs, and
6\% for the high resolution $96^3$ case).
It appears that field saturation is determined by universal behavior
in the partitioning of thermal and magnetic energy,
independent of initial amplitude.
This is true also for the $\Gamma=2$ and $3$
cases, both of which result in mean $1/\beta_B$ profiles similar
to the $\Gamma=5/3$ results shown in Figure \ref{fig:betainv53}.

\begin{figure}
\plotone{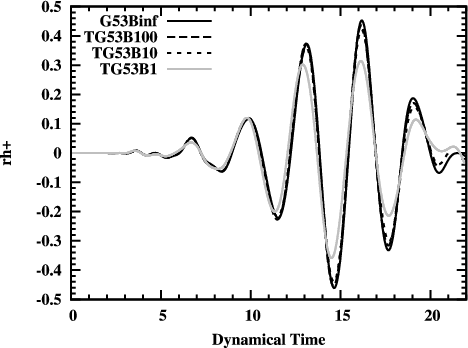}
\caption{Gravitational waveforms ($rh+$) extracted from several
magnetized and unmagnetized $\Gamma=5/3$ cases.
The horizontal time axis is displayed in dynamical code units
$\sqrt{\varpi_E^3/GM}$, and the wave amplitude is 
normalized by $(GM/\varpi_E c^2)^2$.
\label{fig:gw53}}
\end{figure}

\begin{figure}
\plotone{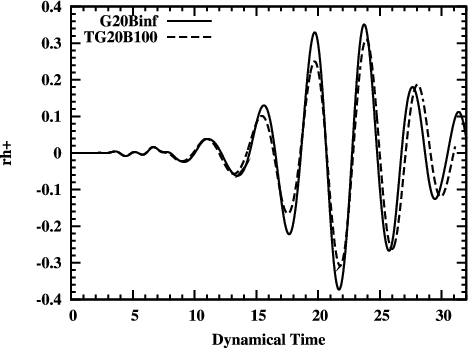}
\caption{Gravitational waveforms with magnetic fields (dashed line)
and without (solid line) for $\Gamma=2$.
\label{fig:gw20}}
\end{figure}

\begin{figure}
\plotone{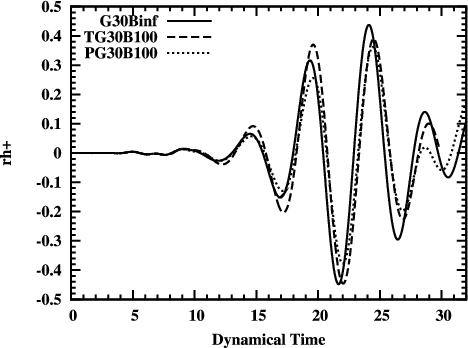}
\caption{Gravitational waveforms with magnetic fields (dashed lines)
and without (solid line) for $\Gamma=3$.
\label{fig:gw30}}
\end{figure}

Consequently, we do not expect gravitational waveforms to be affected
appreciably  by toroidal magnetic fields, as we demonstrate in Figure \ref{fig:gw53}
for the $\Gamma=5/3$ cases,
Figure \ref{fig:gw20} for $\Gamma=2$,
and Figure \ref{fig:gw30} for $\Gamma=3$.
Figures \ref{fig:gw53} through \ref{fig:gw30} plot the quantity $r h+$ normalized by 
the scale factor $(GM/\varpi_E c^2)^2$. Figure \ref{fig:gw30} also includes results
from a poloidal initial field configuration (run PG30B100) for comparison.
Figure \ref{fig:gw53} (corresponding to the $\Gamma=5/3$ cases) closely resembles Figure~9 of \citet{new00}: we match the amplitude and frequency
of oscillations and find that our results are intermediate between the ``D1'' and
``L1'' displays in duration and pattern of the wave signal. It is not until the field
strength is increased to $1/\beta_{B,min} \sim 1$ with locally comparable thermal
and magnetic pressures that we observe amplitude deviations of
order 30\% in Figure \ref{fig:gw53}. 

The first evidence of oscillations in Figure \ref{fig:gw53} occurs at time $t\sim6$
which corresponds to the instant when the $m=2$ mode first begins to dominate
the spectral signal in Figure \ref{fig:mode53}. The global envelope shape (essentially
the overall amplitude) of the gravitational wave emission tracks nicely the growth
and eventual decay of the $m=2$ mode curve in Figure \ref{fig:mode53}. Maximum
peaks in both wave signals and spectral mode profiles correlate precisely at time $t\sim16$,
and both exhibit comparable rise and decay times.
Another point of interest in comparing Figures \ref{fig:gw53} -- \ref{fig:gw30}
is the apparent trend for the start of the wave
signals to be delayed with increasing adiabatic index $\Gamma$ (evident also
in the spectral mode plots). However, we have found that the onset of the
instability is sensitive to a number of numerical factors
(e.g., grid resolution, Courant factor),
and it is difficult to make quantitative conclusions regarding this effect.
For example, the magnetized and unmagnetized $\Gamma=5/3$ high-resolution cases 
resemble Figure \ref{fig:gw53} but for a slight delay of about 1.5
dynamical times, effectively a 10\% temporal shift in the waveform.
However, other aspects of the waveforms are similar between the higher and lower resolution cases: 
the magnetized results are essentially identical to the unmagnetized waveforms, and the
wave amplitudes agree nicely.

Even though the gravitational wave amplitudes are fairly consistent with no obvious correlation
with $\Gamma$ (approximately 0.45, 0.35,
and 0.44 for $\Gamma=$ 5/3, 2 and 3, respectively),
the wavelength of perturbations between the two biggest wave crests
and the burst duration (between leading and trailing wave crests with
amplitude larger than 0.1) do appear to increase monotonically with $\Gamma$.
In particular, we find wavelengths of approximately 3.1$\tilde t$, 4.0$\tilde t$,
and 4.8$\tilde t$ for $\Gamma=$ 5/3, 2, and 3, a fractional increase
(in wavelength) of about 20\%
between $\Gamma=2$ and $\Gamma=5/3$, and about 55\% between $\Gamma=3$ and $\Gamma=5/3$. 
Trends in both amplitudes and wavelengths are consistent with
those of \citet{houser96}, who find the amplitude is independent
of the polytropic index, but $\Gamma=2$ ($\Gamma=3$) fluids generate
gravitational waves with 20\% (58\%) longer wavelengths than $\Gamma=5/3$.

We cannot 
quantify the burst duration as easily since the late time solutions can be inaccurate
due to buildup of numerical errors, and sensitivity to computational
parameters (e.g., Courant factor, artificial viscosity constants). However,
a comparison of Figures \ref{fig:gw53} -- \ref{fig:gw30} clearly shows
a lengthening of the pulse duration by more than 50\% as the models grow stiffer.
Such behavior was observed also by \citet{houser96} and \citet{williams88} in
their unmagnetized studies: stiffer polytropes produce more elongated bars,
rotate more slowly, and undergo more periods of spiral arm ejection and core
recontraction, resulting in longer bursts of gravitational wave signals.
The addition of toroidal magnetic fields to the stellar profile
does not appear to affect these behaviors appreciably as the
burst duration is generally shorter
than the timescale $t_B$ for magnetic braking to take effect:
$t_B \sim \varpi_E/v_A \gtrsim$ 25-50 dynamical (code) units at the field
saturation time for all of the cases we have considered.

\subsection{Poloidal Magnetic Field Configurations}
\label{subsec:polresults}

Unlike the cases which start from initially toroidal magnetic 
field configurations, simulations performed of models which begin
with poloidal fields evolve differently than unmagnetized
cases if the initial field amplitude is large. In particular, we found little
qualitative difference between all the toroidal calculations at both low
($64^3$) and high ($96^3$) grid resolutions, regardless of initial field amplitude. 
For this reason, all of the results presented in Section~\ref{subsec:torresults} 
corresponded to low-resolution cases, with the added advantage of allowing for
many different parameter combinations to be investigated.
However, for the poloidal cases, we observed an increased sensitivity to spatial
resolution as well as a growing impact on bar formation with increasing field strength,
especially for the two largest amplitude cases (PG53B100 and PG53B10).
All poloidal results presented in this section are therefore shown at
high ($96^3$) grid resolution.

\begin{figure}
\plotone{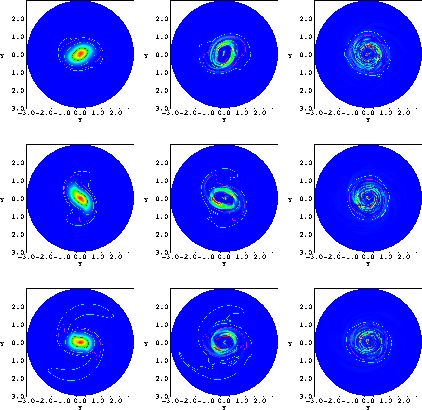}
\caption{Development of the bar-mode
in $\Gamma = 5/3$ simulations with initially poloidal magnetic fields (PG53B100HR, center column, and PG53B10HR, right column)
and without (G53BInfHR, left column).
Rows represent snapshot solutions at different times: $t=$ 15, 17, and 21 for the case without magnetic fields,  $t=$ 17, 19, and 23 for the PG53B100HR case, and $t=$ 15, 17, and 19 for the PG53B10HR case, all
in dynamical (code) units ($\tilde{t} = \sqrt{\varpi_E^3/GM}$).
Contour levels represent mass density in the equatorial plane at four
values normalized to the initial maximum density $\rho_{\text{max,0}}$:
(0.5, 0.05, 0.005, and 0.0009).
Linear color maps represent mass density in the case without magnetic fields
(left column), and magnetic pressure in the cases with fields (center and right columns).
The magnetic pressure color scale is fixed for all the images to
cover the range $0 \rightarrow 10^{-3}$.
\label{fig:pimages}}
\end{figure}

To illustrate the effect of the magnetic field on bar formation, 
Figure~\ref{fig:pimages} shows images of the mass density 
(for the G53BInfHR case) and magnetic pressure 
(for the cases PG53B100HR with $\beta_\mathrm{B,min}=100$, and PG53B10HR 
with $\beta_\mathrm{B,min}=10$), all in the equatorial plane. 
All three sets of images display the same contour levels of the 
mass density (0.5, 0.05, 0.005, and 0.0009), 
normalized to the initial maximum 
mass density $\rho_\mathrm{max,0}$. Images making up the left column for the case 
without a magnetic field are shown at times $t=$15, 17, and 21; images
in the center column are shown for case PG53B100HR at 
times $t=$17, 19, and 23; and images in the right column  are shown for case
PG53B10HR at times $t$=15, 17, and 19, all in dynamical (code) units. 
The PG53B100HR images are shown at later times than those for 
the unmagnetized case to illustrate that a bar forms with a 
shape similar to the no-field case, but delayed approximately two dynamical times. 
For the higher magnetic field strength run, PG53B10HR, there is no indication that 
a structure resembling a bar is going to form at this resolution, 
a point we return to below.

\begin{figure}
\plotone{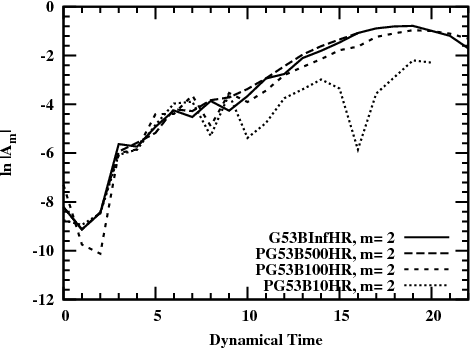}
\caption{Growth of the $m=2$ azimuthal Fourier mode amplitudes 
$|A_m|$ in the mass density, comparing four $\Gamma=5/3$ cases:
unmagnetized (G53BInfHR, dark line)
and magnetized, with initially poloidal magnetic fields (PG53B500HR, long dashed line; 
PG53B100HR, short dashed line; and PG53B10HR, dotted line). 
Results are derived by averaging the density inside a ring centered at 
$\varpi=0.45\varpi_E$ in the equatorial plane and extending one zone
deep in both radial and polar directions.
\label{fig:polmode53-2}}
\end{figure}

\begin{figure}
\plotone{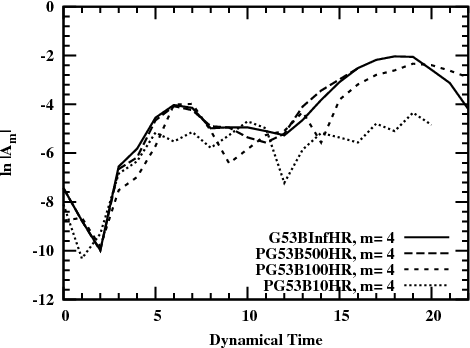}
\caption{Growth of the $m=4$ azimuthal Fourier mode amplitudes 
$|A_m|$ in the mass density, comparing four $\Gamma=5/3$ cases:
unmagnetized (G53BInfHR, dark line)
and magnetized, with initially poloidal magnetic fields (PG53B500HR, long dashed line; 
PG53B100HR, short dashed line; and PG53B10HR, dotted line). 
Results are derived by averaging the density inside a ring centered at 
$\varpi=0.45\varpi_E$ in the equatorial plane and extending one zone
deep in both radial and polar directions.
\label{fig:polmode53-4}}
\end{figure}

The effect of an initially poloidal magnetic field on the growth of the bar mode 
is quantified in the nonaxisymmetric amplitudes $|A_m|$,
computed as described above in Section~\ref{subsec:torresults}. 
In Figures~\ref{fig:polmode53-2} and~\ref{fig:polmode53-4}, we plot the evolution 
of the amplitudes $|A_m|$ of the $m=2$ and $m=4$ modes, respectively, for runs G53BInfHR, PG53B500HR, 
PG53B100HR, and PG53B10HR. It is apparent that for the lowest field case 
(PG53B500HR), the magnetic field has little effect 
on the growth of either mode, but increasing the field amplitude systematically 
suppresses the different modes. For example, the intermediate field case (PG53B100HR)
exhibits growth similar to the no-field case but with a temporal delay
and slightly smaller mode amplitudes, while
modes in the highest field amplitude case (PG53B10HR)
are suppressed by approximately two orders of magnitude relative to the no-field case.
Although not shown, we note that the $m=1$ and $m=3$ modes are similar to those 
observed in the toroidal cases, in that they saturate at levels which are well below the characteristic $m=2$ and 4 profiles of the bar mode.

\begin{figure}
\plotone{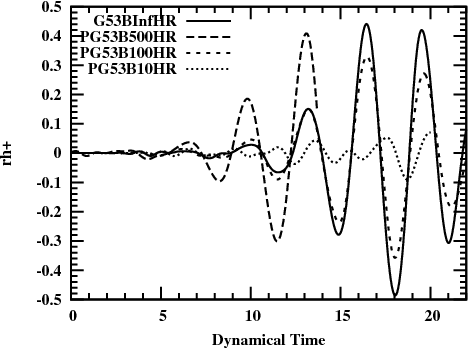}
\caption{Gravitational waveforms ($rh+$) extracted from four
$\Gamma=5/3$ cases: unmagnetized (G53BInfHR, dark line)
and magnetized, with initially poloidal magnetic fields (PG53B500HR, long dashed line; PG53B100HR, short dashed line; and PG53B10HR, dotted line). In this plot, the waveforms produced in the PG53B500HR and PG53B100HR simulations have been shifted in time so that the peaks in the waveforms line up with those of the unmagnetized case. 
\label{fig:polgw53shifted}}
\end{figure}

We expect the gravitational waveforms produced in each of these runs to reflect 
what is seen in the density profiles and the mode amplitude plots. 
In Figure~\ref{fig:polgw53shifted}, we show the quantity $rh_+$ normalized by the 
scale factor $(GM/\varpi_E c^2)^2$, for runs G53BInfHR, PG53B500HR, 
PG53B100HR, and PG53B10HR. For the plot shown, the low-amplitude field case PG53B500HR 
has been shifted back 0.2 dynamical times, and the moderate-amplitude field case PG53B100HR has 
been shifted 1.4 dynamical times so that the maximum of the first large peaks line up. 
As expected, waveforms exhibit systematically smaller amplitudes and greater temporal
delays with increasing initial field amplitude.

\begin{figure}
\plotone{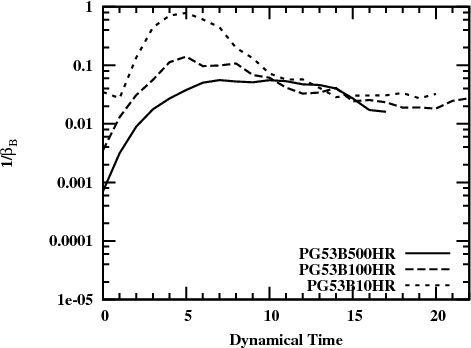}
\caption{Inverse of the magnetic plasma beta (1/$\beta_B$) azimuthally averaged over a ring in the equatorial plane at the radius $0.65\varpi_E$.
}
\label{fig:polshellbetainv53}
\end{figure}

In Figure~\ref{fig:polshellbetainv53}, we plot the inverse plasma beta ($1/\beta_B$), azimuthally 
averaged over the ring at $\varpi=0.65 \varpi_E$ in the equatorial plane, for 
runs PG53B500HR, PG53B100HR, and PG53B10HR.
For the lowest field case (PG53B500HR), in which the dynamics are affected only slightly, 
the magnetic pressure is never more than about 4\% of the fluid pressure in this ring. 
However, for the highest field case (PG53B10HR), in which the bar mode is completely 
suppressed, the magnetic pressure is already just below 4\% of the fluid pressure in the 
initial configuration, and peaks at around 80\% of the fluid pressure within five dynamical times
before leveling off again at $\sim$4\%.
Under these conditions, we would expect the magnetic field to have a significant effect 
on the overall evolution of the star.

\begin{figure}
\plotone{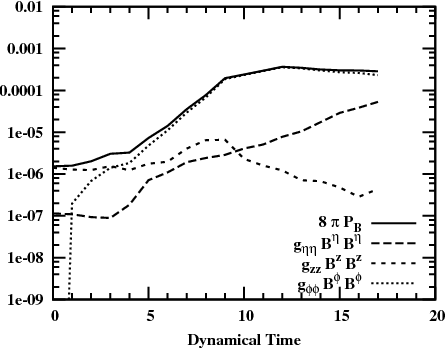}
\plotone{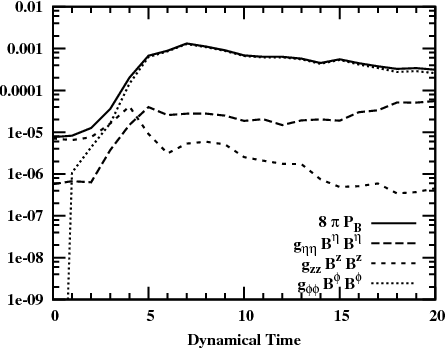}
\plotone{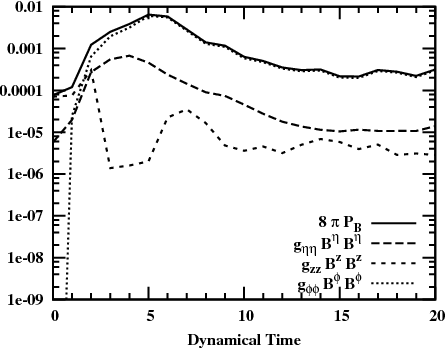}
\caption{We show $8\pi$ times the magnetic pressure, as well as the contributions to this quantity from each of the three cylindrical components of the magnetic field, all azimuthally averaged over the ring at $\varpi=0.65 \varpi_E$ in the equatorial plane, for runs PG53B500HR (top), PG53B100HR (center), and PG53B10HR (bottom).}
\label{fig:polshellmagpress}
\end{figure}

\begin{figure}
\plotone{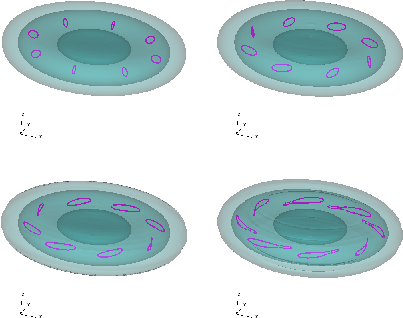}
\caption{Magnetic field lines for the PG53B100HR simulation, at times 0 and 1 $\tilde{t}$ (top row), and 2 and 3 $\tilde{t}$ (bottom row). Also shown are mass density contours at three values, normalized to the initial maximum density $\rho_\mathrm{max,0}$: (0.5, 0.05, 0.005).}
\label{fig:polblines}
\end{figure}

We can understand the immediate onset of the field amplification 
observed in Figure~\ref{fig:polshellbetainv53}
by considering the evolution of individual components of the magnetic field.
In Figure~\ref{fig:polshellmagpress}, we plot $8\pi$ times the magnetic pressure, 
as well as the contributions to this quantity from each of the three 
cylindrical components of the magnetic field, all azimuthally averaged over 
the ring at $\varpi=0.65 \varpi_E$ in the equatorial plane, for runs 
PG53B500HR, PG53B100HR, and PG53B10HR. For all three cases, 
we see that the increase in the overall magnetic pressure tracks the increase 
in the azimuthal contribution to that pressure. This behavior, in addition to 
the immediate field amplification, is characteristic of the $\Omega$-dynamo, 
which one would expect to be active with a poloidal magnetic field present. 
This effect can be seen qualitatively in Figure~\ref{fig:polblines}, in which 
we show magnetic field lines at early times for the PG53B100HR case. 
The magnetic field lines, as well as mass density contours, are shown at 
times $t=0, 1, 2$, and $3\tilde{t}$. The field lines show immediate and dramatic 
stretching in the azimuthal direction, contributing to the field amplification
observed in Figures~\ref{fig:polshellbetainv53} and~\ref{fig:polshellmagpress}.

Another important point from Figure \ref{fig:polshellbetainv53} is that the 
late time saturation level of $1/\beta_{B}$ is approximately the same in all 
three simulations. A similar behavior and level of saturation are evident in 
Figure \ref{fig:betainv53} for the toroidal cases. This remarkable independence 
from the initial field strength and orientation is important, as it indicates that
a common and robust physical mechanism is at play setting the final saturation level. 
Also, we see from Figure~\ref{fig:polshellmagpress}
that the final magnitude of the azimuthal component ($g_{\phi\phi}B^\phi B^\phi$) 
is approximately the same in all three simulations, and the second most 
significant component, the radial one ($g_{\eta\eta}B^\eta B^\eta$), always 
finishes within an order of magnitude of the azimuthal component. This suggests that 
these two components have either achieved a static final configuration common to 
all the simulations, or they are feeding off of one another and exchanging energy 
with the gas through some equilibrium process. The latter is consistent with
expectations of the MRI, which gradually takes over as the $\Omega$-dynamo saturates,
and which is expected to be active in all of our stellar models according to
criterion (Equation~(\ref{eqn:mricondition})).

We end this section with a few words of caution regarding the convergence of results
for those cases run with poloidal initial fields. The smallest B500 case is reasonably
converged between low ($64^3$) and high ($96^3$) grid resolutions, resulting
in nearly identical gravitational wave frequencies and maximum peak differences of
about 25\% in amplitude. However, the intermediate B100 case exhibited vastly different 
solutions at low and high resolutions. The bar mode was essentially suppressed completely
at low resolution, but managed to form nicely at $96^3$ zones, demonstrating the
sensitive and demanding nature of these calculations. We cannot therefore be certain 
that the largest amplitude (B10) case is fully converged and does not form a bar.
It would require even larger computational grids to confirm this result, something
that is beyond our current allocation resources. At issue is the resolution of
the winding of the field lines as the star rotates, the accurate capturing of reconnection
events, and mostly the severe restrictions placed on the stability time step from
magnetically dominated MHD and fast Alfv\'en velocities as the field works its way
through the stellar atmosphere.

\section{Conclusions}
\label{sec:conclusion}

We have studied the growth of the dynamical bar-mode instability in 
differentially rotating magnetized neutron stars through a set of 
numerical Newtonian MHD calculations. The calculations explored both 
toroidal and poloidal initial field distributions of differing strengths, 
as well as the role of the equation of state.

For our initially toroidal field configurations, field amplification 
always saturated at a level insufficient  
to strongly affect the dynamics of the bar mode. Even the most extreme 
case (TG53B1, $\beta_\mathrm{B,min}=1$), with equal initial magnetic
and hydrodynamic pressures within the toroidal loop, gave only a $\sim30$\% 
change in gravitational wave properties. 
Otherwise evolution proceeded quite similarly to our unmagnetized reference cases.

The effects were larger  
for the poloidal configurations. In the most extreme case considered (model
PG53B10HR with $\beta_\mathrm{B,min}=10$), the magnetic field was sufficient to
completely suppress the formation of a bar. However, in that case, we started  
with the magnetic field already within a factor of $\sim$30 of being  
dynamically more important than thermal pressure in some portions of the star.
It then only took a few dynamical times of field growth for $\beta_B$ to  
approach unity, a timescale that is much shorter than the bar deformation time
for azimuthal Fourier modes to reach appreciable amplitudes.  
Also, due to the computationally demanding nature of evolving strong poloidal fields,
we cannot verify that this result will not change with increased grid resolution.
For less extreme initial conditions, 
the effects of including poloidal field components were consistently more modest
with decreasing initial field amplitude. These lower amplitude calculations are also
less demanding computationally and can, like the toroidal cases, be more easily 
checked for convergence.
The principle effects of introducing poloidal fields are in systematically
delaying the onset of the bar mode and in suppressing both
Fourier mode and gravitational wave amplitudes, all of which become negligible with
initial field amplitudes below the threshold of $1/\beta_\mathrm{B,min} \lesssim 10^{-2}$.

Overall, our results suggest that the effect of magnetic fields on the 
emergence of the bar-mode instability in neutron stars is not likely to be 
very significant.
Particularly considering that collapse progenitor models predict realistic 
field configurations that are dominantly toroidal in nature with
toroidal and poloidal components of order
$10^{10}$G and $10^6$G, respectively \citep{heger05}, substantially below most of the 
field configurations we have considered.
Thus, except in special cases where neutron stars are 
born very highly magnetized, we might still expect them to be good 
gravitational wave sources if their rotational kinetic energies 
exceed the critical bar-mode instability parameter.

\acknowledgments

Computations were performed at the Barcelona Supercomputing Center (BSC)
under activity AECT-2007-3-0002, the
Lawrence Livermore National Laboratory (LLNL), 
the College of Charleston (CoC), and at the High Performance Academic 
Computing Environment at Washburn University (HiPACE). 
This work was performed in part under the auspices of the 
U.S. Department of Energy by Lawrence Livermore National Laboratory
under contract no. DE-AC52-07NA27344. P.C.F. gratefully acknowledges the support of the College of Charleston 4th Century Initiative and the South Carolina Space Grant Consortium. J.A.F. acknowledges financial
support from the Spanish Ministry of Education and Science (AYA 2007-67626-C03-01).


\end{document}